\documentclass[showpacs, prb,twocolumn,preprintnumbers , superscriptaddress, aps]{revtex4-2}
\usepackage[dvips]{graphics}
\usepackage{color}
\definecolor{dred}{rgb}{0,0,0.6}
\usepackage{amssymb}   
\usepackage{amsmath}
\usepackage{braket}
\usepackage{upgreek}
\usepackage[mathscr]{euscript}
\usepackage{color}
\usepackage{float}
\usepackage[toc,page]{appendix}
\usepackage[normalem]{ulem}

\DeclareUnicodeCharacter{2212}{-}

\usepackage{tikz}
\usepackage{bbold}
\usepackage{makecell}
\usepackage{pifont}   
\usepackage{graphicx} 
\usepackage{dcolumn}  
\usepackage{bm}       
\usepackage{multirow} 
\usepackage{placeins}
\usepackage[colorlinks]{hyperref}
\usepackage{mathtools}

\usepackage[normalem]{ulem}

\usepackage{soul} 
\usepackage{ulem} 

\usepackage{mathrsfs}

\usepackage{multirow}
\usepackage[margin=1in]{geometry}
\usepackage{subcaption}

\usepackage{ragged2e}
\DeclareCaptionFormat{justified}{\justifying#1#2#3}
\begin{document}

\title{Floquet-Induced Valley Polarization and Topological Phases in Twisted Bilayer Graphene Heterostructures}
\author{A. Belayadi}
\email{abelayadi@usthb.dz}
\affiliation{Department of theoretical physics$,$ University of Science and Technology Houari Boumediene$,$ Bab$-$Ezzouar$,$ Algeria.}

\author{I. A. Assi}
\affiliation{Department of Physics and Physical Oceanography$,$
Memorial University of Newfoundland and Labrador$,$ St. John’s$,$ Newfoundland $\&$ Labrador$,$ Canada A1B 3X7}

\author{A. Johansson}
\email{annika.johansson@mpi-halle.mpg.de}
\affiliation{ Institute of Physics, Martin Luther University Halle-Wittenberg, 06099 Halle, Germany.}

\author{C. I. Osuala }
\affiliation{Department of Physics and Astronomy, Ohio University, Athens, Ohio 45701, USA}

\author{H. Bahlouli}
\affiliation{Physics Department$,$ King Fahd University of Petroleum $\&$ Minerals$,$ Dhahran 31261$,$ Saudi Arabia}
\affiliation{Interdisciplinary Research Center for Advanced Materials, Saudi Arabia.}

\author{Udo Schwingenschl\"ogl}
\email{udo.schwingenschlogl@kaust.edu.sa}
\affiliation{Physical Science and Engineering Division, King Abdullah University of Science and Technology, Thuwal 23955-6900, Saudi Arabia}

\author{J. P. F. LeBlanc}
\affiliation{Department of Physics and Physical Oceanography$,$
Memorial University of Newfoundland and Labrador$,$ St. John’s$,$ Newfoundland $\&$ Labrador$,$ Canada A1B 3X7}
\affiliation{Compute Everything Technologies Ltd.$,$ St. John's$,$ Newfoundland $\&$ Labrador$,$ Canada}

\author{A. Abbout}
\affiliation{Physics Department$,$ King Fahd University of Petroleum $\&$ Minerals$,$ Dhahran 31261$,$ Saudi Arabia}
\affiliation{Interdisciplinary Research Center for Advanced Quantum Computing, Saudi Arabia.}
\date{\today}

\begin{abstract}
We theoretically investigate twisted bilayer graphene (tTBG) heterostructures subjected to circularly polarized light. Inversion symmetry is broken through the combined application of an out-of-plane electric field and the incorporation of insulating substrates, enabling precise control of the valley degrees of freedom and associated topological phases. We identify a regime in which the valley degeneracy is lifted: one valley retains a gapless, linear Dirac-like dispersion, whereas the other develops a finite band gap and becomes electronically inactive. The interplay between Floquet driving and inversion-symmetry breaking induces topological phase transitions characterized by Chern numbers (C=-2), (C=-1), and (C=0), together with the emergence of perfectly valley-polarized domain-wall channels. Furthermore, magnetic asymmetry arising from proximity to adjacent two-dimensional ferromagnetic contacts enables spin–valley filtering. These findings establish the potential of graphene-based heterostructures as highly tunable systems for controlling spin and valley degrees of freedom.
\end{abstract}

\maketitle

\section{Introduction} 
The ability to control the valley degree of freedom in two-dimensional materials is central to the development of valleytronic devices. Several approaches have been explored, including strain engineering \cite{yu2024manipulating,shayeganfar2022strain,Hadadi_2023,Hadadi_2025,Belayadi2025valley}; optical methods for controlling and detecting valley polarization \cite{ho2020measuring,soni2025enhanced}; and sublattice-symmetry breaking in graphene-based heterostructures \cite{belayadi2023valley,goh2023valleytronics}. More generally, the controlled breaking of spatial-inversion and time-reversal symmetries provides an effective route for modifying the electronic, valley, and topological properties of quantum materials \cite{hughes2011inversion,qi2008topological,fu2007topological,moore2007topological}. Stacking two-dimensional crystals with compatible or complementary symmetries produces van der Waals heterostructures whose electronic properties can differ substantially from those of the constituent layers and can be further modified by the relative stacking configuration and interlayer twist angle \cite{du2021engineering,kang2017layer,lim2014stacking} \cite{fox2023stacking,lim2014stacking}.

Topological materials exhibit robust transport properties arising from gapless edge modes, making them promising platforms for quantum information applications. In addition to static structural engineering, time-dependent electromagnetic fields provide a dynamical route for accessing electronic states that are absent in equilibrium. When a material is subjected to a periodic drive, its electronic structure is described in terms of Floquet states and quasienergy bands \cite{muliana2011time}. These driven states can be probed using techniques such as time- and angle-resolved photoemission spectroscopy  \cite{ARPES_Floquet_Wang2013,ARPES_Floquet_Mahmood2016}. In particular, circularly polarized light breaks time-reversal symmetry because time reversal reverses the helicity of the optical field. As a result, irradiation can generate nontrivial Floquet phases, including Floquet topological-insulator and quantum-anomalous-Hall regimes \cite{Floquet_top_ins_Chou2020,QAHE_McIver2020}. 

Within the off-resonant, high-frequency regime, the driven system can often be represented by an effective time-independent Hamiltonian. In this description, circularly polarized light generates a Haldane-like mass term whose sign depends on the light helicity \cite{haldane1988model,Floquet_Eckardt_2015,Floquet_Ibsal_2021}. The Haldane model was originally introduced for a honeycomb lattice and has since been investigated in ultracold atoms \cite{cheng2019topological}, bilayer graphene \cite{qu2017valley,dal2017one,luo2021tuning,zhai2020spin}, quantum materials \cite{pan2014valley,zhou2017valley}, and photonic systems \cite{prudencio2021first}. For spinless fermions at half filling, a nonzero Chern number is associated with a gapped bulk and conducting boundary states, in analogy with the quantum Hall effect \cite{liu2023research}. This mechanism is particularly relevant to graphene-based moiré systems, where light-induced masses can compete with interlayer coupling, electrostatic potentials, and substrate-induced sublattice asymmetry.

Electronic properties in graphene multilayers are also strongly influenced by van der Waals interactions between adjacent layers \cite{lu2014van,klein2001electronic}. This influence is especially pronounced in twisted bilayer graphene (tBLG), where the relative rotation of the two graphene sheets produces a moiré structure and modifies the low-energy band dispersion \cite{hou2025strain,nimbalkar2020opportunities,cai2021fabrication}. Substrates can provide an additional source of symmetry breaking. For example, coupling graphene to SiC can induce a sublattice-dependent potential, thereby breaking sublattice symmetry and modifying the low-energy Dirac states of the graphene layers \cite{wang2016gaps,nevius2015semiconducting,giovannetti2007substrate,zhou2007substrate}. Depending on the substrate arrangement and the applied displacement field, these potentials can generate layer- and valley-dependent energy shifts and can alter the opening or closing of bulk gaps.

Magnetic proximity provides a complementary means of controlling the internal degrees of freedom. Coupling graphene to a magnetic material can induce exchange splitting, spin polarization, and, in suitable structures, enhanced spin–orbit effects \cite{zollner2020swapping,zhai2022proximity}. Such proximity effects have been demonstrated or proposed in graphene-based heterostructures \cite{gerber2025tunable,asim2025high,zollner2021bilayer}. An externally applied displacement field can provide electrostatic control over the resulting spin- and valley-dependent band structure. Combining magnetic proximity with substrate-induced sublattice asymmetry is therefore a promising strategy for realizing electrically tunable spin–valley filtering and topological transport.

\begin{figure}[tp]
\centering
\includegraphics[width=8cm, height=10cm]{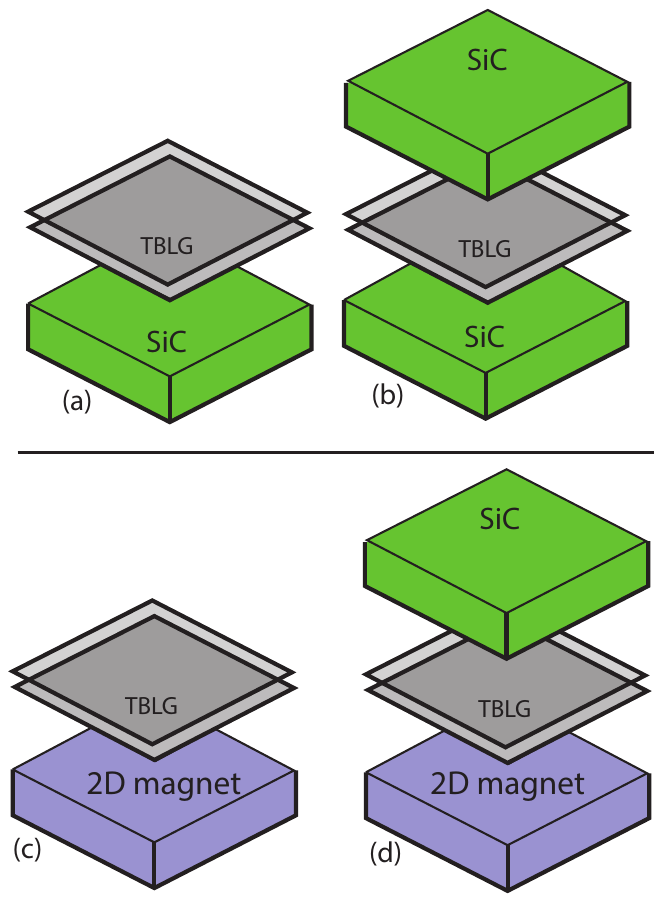}
\vspace{-0.15cm}
\caption{Scheme of a tBLG-based heterostructure encapsulated by SiC substrate(fig. (a), (b)) or ferromagnetic (like Cr$_2$Ge$_2$Te$_6$) contact (fig. (c), (d)) which induce, respectively, sublattice imbalance  or exchange interactions in the active layer defined by the substrate nature. Several configurations are  considered  such as bottom, up or bottom-up approach to vdW engineering (a, b, c and d)}
\label{fig1}
\end{figure}

Here, we investigate the combined effects of circularly polarized light, substrate-induced sublattice asymmetry, interlayer bias, and magnetic proximity on the electronic structure and topological phases of tBLG heterostructures. These perturbations modify the valley- and layer-resolved mass terms and can open, close, or invert bulk gaps, thereby driving transitions between distinct topological phases. We consider two classes of structures, as illustrated schematically in Fig. \ref{fig1}: (i) tBLG on SiC substrate (tBLG/SiC) and SiC-encapsulated tBLG, which are used to study substrate-induced sublattice asymmetry; and (ii) tBLG coupled to a magnetic layer such as Cr$_2$Ge$_2$Te$_6$ in tBLG/CGT and SiC/tBLG/CGT  configurations, which are used to study exchange proximity effects \cite{zollner2021data,zollner2020swapping,zollner2022proximity, zollner2023erratum}. Other magnetic materials, including EuO and yttrium iron garnet (YIG), may provide alternative sources of magnetic proximity in related graphene-based spintronic structures \cite{singh2017strong,su2017effect,hallal2017tailoring,yang2013proximity}.

In the presence of light, bias voltage, sublattice asymmetry, and/or exchange coupling, we find that the phase diagram of tBLG heterostructures host a rich variety of topological phases. Specifically, we show that tuning the interlayer bias at fixed substrate-induced asymmetry can produce a valley-selective regime in which one valley retains a gapless, approximately linear Dirac-like dispersion while the other is gapped. We further examine interfaces between regions with different proximity-induced masses and topological indices, where localized channels with strongly valley-selective character can emerge. Finally, by incorporating magnetic proximity, we demonstrate a route toward electrically controlled spin–valley filtering. Together, these results establish tBLG heterostructures as a versatile platform for studying the interplay between Floquet engineering, valley polarization, and proximity-induced topological phases.

The remainder of this paper is organized as follows. In Sec. II, we introduce the low energy tight-binding model that describes our twisted bilayer graphene along with Haldane term that describes interaction with circularly polarized light and the interaction with substrate, this full Hamiltonian is used to investigate the electronic and topological properties of the system. In Sec. III, we present and discuss our numerical results, focusing on electrically tunable topological phases, valley-polarized interface states, and spin–valley filtering induced by magnetic proximity effects. Finally, Sec. IV summarizes our main findings and conclusions.

\section{Methodology}

{\bf Tight-binding model.} 
The tight binding Hamiltonian, for such a tBLG  vdW structure takes the form $H = H_{\rm tBLG} + H_{h} + H_{S}$ , where $H_{\rm tBLG}$ describes the tBLG bilayer Hamiltonian, $ H_{h} $ the Haldane Hamiltonian, and $H_{S}$ accounts for the additional substrate-induced interaction terms. 

We describe the twisted bilayer graphene (tBLG) using a single-orbital tight-binding model based on the carbon $p_z$ orbitals. The Hamiltonian is given by
\begin{equation}\label{eqn1}
   H_{\rm tBLG}=\sum_{i, \sigma}^N\left( \epsilon_i+V \tau_z \right) { c}_{i\sigma}^\dagger { c}_{i\sigma} +\sum_{<i, j>,\sigma} t_{i, j}\left(\mathbf{d}_{i j}\right) { c}_{i\sigma}^\dagger { c}_{j\sigma} ,
\end{equation}
where ${ c}_{i\sigma}^\dagger ({ c}_{i\sigma} )$ is the creation (annihilation) operator for an electron at site $i$ with spin $\sigma \in \{\uparrow, \downarrow\}$, on-site $\epsilon_i$ and $V$ is a bias voltage term whereas $\tau_z$ is defined in such a way that $\tau_z=+1$ ($\tau_z=-1$)  if $i$ belongs to sites in the top (buttom) layer.

The term $t_{ij}$ defines the intralayer and interlayer transfer integrals defining the hopping between the sites $i$ and $j$. 
Due to twisting, we define the interatomic coupling to be exponentially decaying away from their centers \cite{bandeira2025energy,duarte2024moire,suarez2010flat}. For this reason, the hopping strength between two neighboring sites decays exponentially as a function of their separation distance $\left| \mathbf{d}_{\mathbf{i j}} \right|$. Within Slater-Koster formalism  \cite{koshino2015electronic,moon2013optical}, the hopping integral $t_{ij}$  is defined in terms of $\sigma^{-}$ and $\pi^{-}$-like bonds and is expressed in terms of the transfer integrals of the $pz$-orbitals

\begin{equation}\label{eqn2}
t_{i j}=-V_{p p \pi}\left[1-\left(\frac{\mathbf{d}_{\mathbf{i j}} \cdot \mathbf{e}_{\mathbf{z}}}{d_{i j}}\right)^2\right]-V_{p p \sigma}\left(\frac{\mathbf{d}_{\mathbf{i j}} \cdot \mathbf{e}_{\mathbf{z}}}{d_{i j}}\right)^2,
\end{equation}
where
\begin{equation}
V_{p p \pi}=V_{p p \pi}^0 \exp \left[-\left(\frac{d_{i j}-a_{c c}}{\delta_0}\right)\right],
\end{equation}
and
\begin{equation}
     V_{p p \sigma}=V_{p p \sigma}^0 \exp \left[-\left(\frac{d_{i j}-d_0}{\delta_0}\right)\right],
\end{equation}
with $V_{pp\pi}^0= -2.7 \mathrm{~eV} $ and $ V_{pp\sigma}^0 \simeq 0.48 \mathrm{~eV}$. Within the in plane region, we consider only hopping interactions within the first nearest neighbor approximation. For interlayer interactions, only hopping energies for $d \le 4 a_{cc}$ are considered due to the exponential decay of the hopping matrix elements  \cite{mirzakhani2020circular,mirzakhani2025magnetism}. The decay length is $ \delta _0= 0.184 a_{cc}$, where $a_{cc}$ is the carbon-carbon distance of graphene and $\mathbf{e}_{\mathbf{z}}$ is a unit vector perpendicular to the graphene sheets, also called the out-of-plane direction. 

The second term in our model Hamiltonian is the Haldane mass term, which  describes the effect of irradiating tBLG with circularly polarized light \cite{cheng2019topological,luo2021tuning}, 
\begin{equation}
    H_{h}=i \frac{\lambda_h}{3 \sqrt{3}} \sum_{\langle i, j \rangle,\sigma} v_{i j} c_{i \sigma}^{\dagger} c_{j \sigma}+H.c,
\end{equation}
where $\lambda_h$ is the strength of the coupling and $v_{i j}$ defines the orientation of the hopping terms, either clockwise (+1) or counter-clockwise (−1). In the high-frequency range $10^{13}-10^{14} \mathrm{~Hz}$, $\lambda_h$ is of order of tens of $\mathrm{meV}$  \cite{zhai2020spin,kitagawa2011transport,ezawa2013photoinduced}.

Finally, the effect induced by the substrate is
\begin{equation}\label{eqn3}
\begin{aligned}
H_S= & \Delta_z \mathcal{P}_z  \sum_{i, \sigma} \xi_i   c_{i \sigma}^{\dagger} c_{i \sigma} \\
               & +\mathcal{P}_z \sum_{i, \sigma, \sigma^{\prime}} \lambda_m^i c_{i \sigma}^{\dagger} c_{i \sigma^{\prime}}[\hat{\mathbf{m}} \cdot \hat{\mathbf{s}}]_{\sigma \sigma^{\prime}}
\end{aligned}
\end{equation}
where $\mathcal{P}_z$ projects onto the top (bottom) for $z=1$ ($z=-1$) graphene sheet depending on the substrate location. For instance, $\mathcal{P}_{z=-1}$, i.e., the creation (annihilation) operator ${\bf c}_{i\sigma}^\dagger ({\bf c}_{i\sigma} )$ acts only on the sites $A_1$ and $B_1$, hence $\mathcal{P}_{-1}=\operatorname{diag}\left(\mathbb{1}_{2 \times 2}, \mathbb{0}_{2 \times 2}\right)$ in the lower layer sublattice basis. Similarly, if $\mathcal{P}_{z=\pm 1}$, i.e. it acts on both layers sublattice basis with $\mathcal{P}_{\pm1}=\operatorname{diag}\left(\mathbb{1}_{2 \times 2}, \mathbb{1}_{2 \times 2}\right)$ is used in the above expression of $H_{S}$. Furthermore, $\Delta_z$ is the staggered on-site potential for a given layer, 
where $\xi_i=\pm 1$ defines the sublattice asymmetry  due to different local environments experienced by the two sublattices in the heterostructure. The final term models magnetic systems by introducing a sublattice-dependent exchange coupling $\lambda_m$. This term breaks time-reversal symmetry and induces a spin magnetization that is oriented along the $z$-axis.

The topological phase is characterized by the charge Chern number  \cite{pan2014valley,tse2011quantum} 

\begin{equation}\label{eqn4}
\begin{aligned}
C=\frac{1}{2 \pi} \sum_n \int_{\mathrm{BZ}} d^2 \boldsymbol{k} \Omega_n(\boldsymbol{k})
\end{aligned}
\end{equation}

where $\Omega_n$ is the momentum-space Berry curvature  \cite{ren2016topological}. In the out-of-plane direction for the $n$-th sub-band, we have

\begin{equation}\label{eqn5}
\begin{aligned}
\Omega_n(\boldsymbol{k})=-\sum_{n^{\prime} \neq n} \frac{2 \operatorname{Im}\left\langle\psi_{n k}\right| v_x\left|\psi_{n^{\prime} k}\right\rangle\left\langle\psi_{n^{\prime} k}\right| v_y\left|\psi_{n k}\right\rangle}{\left(\varepsilon_{n^{\prime}}-\varepsilon_n\right)^2} .
\end{aligned}
\end{equation}

The summation is over all occupied valence bands $\varepsilon_{n}$ in the first Brilloin zone below the bulk band gap, $\psi_{nk}$ is the Bloch state, and $v_{x(y)}$ is the velocity operator along the $x(y)$ direction. For more details, we refer to \cite{fukui_2005} 

Since we are working in the tight-binding framework, a topological insulator state is labeled at least by one nonzero topological invariant. 


\section{Results}
In this work, we present a comprehensive theoretical investigation of the electronic properties of a twisted bilayer graphene (tBLG) heterostructure irradiated by a circularly polarized light (CPL). Our central goal is to show valley tunability and control of topological phases achieved by introducing various forms of structural and electronic asymmetries induced by different substrates.  For instance, (i) Inversion symmetry is explicitly broken by introducing sublattice and/or interlayer asymmetry. This can be achieved either by applying an out-of-plane electric field (bias voltage) or by placing the structure in proximity to an insulating substrate, such as h-BN or SiC, which induces an asymmetric potential in one or both BLG layers.  (ii)Introducing magnetic asymmetry via magnetic proximity effects from a neighboring two-dimensional magnetic material. Such asymmetry is crucial for the emergence of topological phases, including the Quantum Anomalous Hall (QAH) and Quantum Spin Anomalous Hall (QSAH) states.
    
To this end, we first present the band structure of tBLG, in the absence of irradiation, considering a bias voltage and/or sublattice asymmetry breaking that arises from the graphene–substrate interaction being active in one or both layers. The electronic band structure of tBLG exhibits different features compared to BLG. 

\begin{figure}[tp]
\centering
\includegraphics[width=8cm, height=8cm]{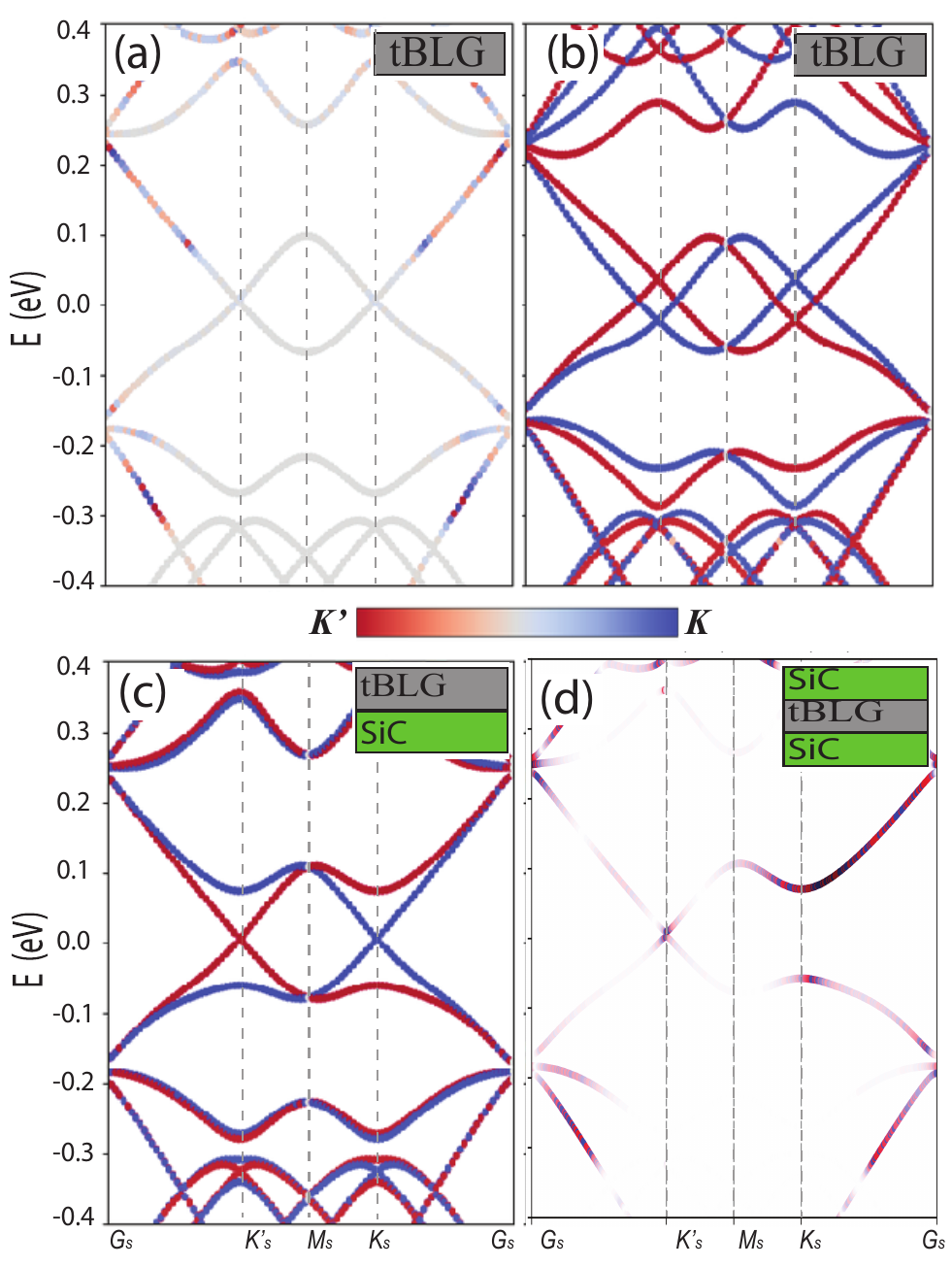}
\vspace{-0.15cm}
\caption{Valley-resolved band structure of tBLG in the absence of irradiation. Panels (a) and (b) are with ($V=50 \ meV$) and without ($V=0 \ meV$) bias voltage, respectively. Band structures of tBLG with broken sublattice symmetry, $\Delta=200 \ meV$, in one (panel (c)) and both layers (panel (d)). The mixed colors is due to valley degeneracy.)} 
\label{fig2}
\end{figure}


Figure \ref{fig2} shows the valley-projected band structures for different symmetry-breaking scenarios. In all cases considered here, the system remains globally gapless. In the absence of both a bias voltage and sublattice symmetry breaking, the band structure exhibits two degenerate Dirac cones associated with the two valleys, as shown in Fig.~\ref{fig2}(a). Applying an out-of-plane bias voltage lifts the valley degeneracy but does not open a gap, leaving the system gapless with overlapping energy ranges between the valley-polarized bands (Fig.~\ref{fig2}(b)). When sublattice symmetry is broken in only one layer, for example in a tBLG-SiC heterostructure, particle-hole symmetry is broken and a gap opens at a single Dirac cone of each valley-polarized state. Nevertheless, a global band gap is not established, and the system remains gapless overall (Fig.~\ref{fig2}(c)). Finally, when sublattice symmetry breaking is present in both layers, as in a SiC-tBLG-SiC structure, the symmetry between the top and bottom layers restores the valley degeneracy. As a result, no global gap emerges and the system remains gapless despite the presence of substrate-induced potentials. In this case, 
the valley lifting of the bands is lost, also resulting in a gapless state

On the basis of these statements, inducing symmetry breaking through the substrate and by tuning the bias voltage could lift the valley degeneracy, leading to distinct energy ranges for the two valleys. The crucial step is then opening a full band gap across the system, which should yield interesting valley-dependent properties. This gap opening can be achieved by combining the existing symmetry breaking and bias voltage with circularly polarized light irradiation.

\subsection{Electronic and Topological Effects in a Layer-Polarized Heterostructure with Sublattice Asymmetry}

We investigate a twisted bilayer graphene heterostructure subjected to circularly polarized light in the presence of sublattice asymmetry. To obtain material-independent topological phase diagrams, we fixed the parameter subjected to circularly polarized light in the presence of sublattice asymmetry. To obtain material-independent topological phase diagrams, we fixed the parameter $\lambda_h $ at its experimentally relevant value of 10 meV  \cite{kitagawa2011transport, ezawa2013photoinduced}.  Our analysis then focuses on the interplay between the staggered sublattice potential $\Delta$ , 
applied to the bottom layer and varied within the range $|\Delta| \leq 200$ meV, and the interlayer bias $V$, with $|V| \leq 170$ meV, consistent with experimentally accessible values \cite{velasco2012quantum, zhang2009direct}.  

\begin{figure}[tp]
\centering
\includegraphics[width=8cm, height=18cm]{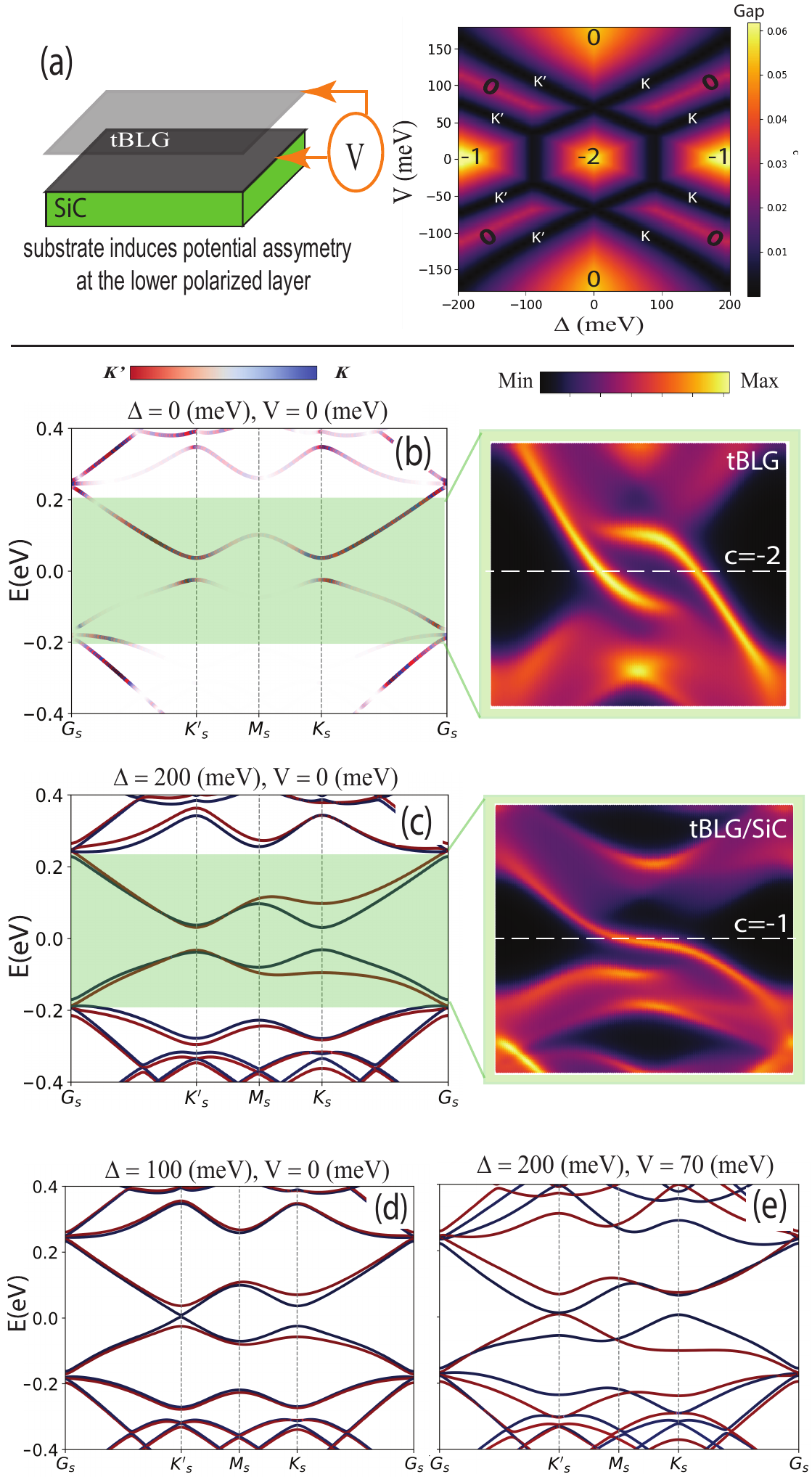}
\vspace{-0.15cm}
\caption{In panel (a), we show the sketch of the adopted tBLG-based heterostructure
 and we present the bulk band gap and the corresponding Chern number as a function of the external tuning parameters, in the presence of irradiation. In (b) and (c) we plot the gaped valley-resolved band structure and its surface spectral function for  $\Delta=100~\text{meV}$ and  $\Delta=200~\text{meV}$.  Panels (d) and (e)   are for  $\left(  \Delta=100~\text{meV},  V=0~\text{mV}\right)$ and  $\left(\Delta=200~\text{meV},    V=70~\text{mV}\right)$. 
 The light-green boxes (right panels), in figures (b) and (c),  show the edge states that are obtained for a system finite in the $y$-direction, and therefore possesses edges that can possibly host one or more topological states.}
\label{fig3}
\end{figure}

In  Fig. \ref{fig3}(a) , we show the bulk band gap and the corresponding Chern number as a function of the external tuning parameters. By tuning the bias voltage, the gap closes and reopens then the Chern number switches from $C=-2$ to $C=-1$ or $C=0$, depending on the strength of staggered sublattice potential $\Delta$ as indicated. Increasing the bias voltage drives a sequence of topological phase transitions marked by bulk-gap closing and reopening events. For small values of $|\Delta|$, the Chern number changes from $C=-2$ to $C=-1$, whereas for larger $|\Delta|$ it evolves from $C=-1$ to $C=0$.  The key to accessing these topological phases lies in the ability to electrically control the band topology. The phase boundaries, highlighted by the black region, indicate points where the direct band gap closes at a specific valley labeled by K and K$'$. 

The combined effect of the substrate-induced staggered potential and the applied bias voltage breaks inversion symmetry and lifts the valley degeneracy.   
By continuously tuning the bias voltage, we observed that the band gap closed and subsequently reopened, signaling a transition to a new topological state, as indicated by the corresponding Chern number.

The application of CPL 
fundamentally alters the band structure by dynamically breaking time-reversal  symmetry. This allows for the stabilization of states with a non-zero Chern number $(C=-2)$   at  $\left| V\right|$ and $\left| \Delta  \right|$ smaller than 50 meV.


In the absence of staggered sublattice potentials and interlayer bias ($\Delta=0$ and $V=0$), circularly polarized light opens a topological gap of identical magnitude in both valleys. Because particle-hole symmetry is preserved, the valley-resolved band structures remain symmetric. The photo-induced gap drives the system into a Photo-induced Quantum Hall State (PQHS) with a total Chern number (C=-2). Consistent with the bulk-boundary correspondence, this phase hosts two chiral edge states in a finite-width geometry, as shown in Fig.\ref{fig3}(b).

In what follows, we assumed a surrounding medium that induced a staggered potential in tBLG, with the corresponding values  
$\left| \Delta \right|=  100  $ (meV) and $ \left| \Delta \right|=  200  $ (meV). These symmetry-breaking parameters can be experimentally tuned by adjusting the buffer layer or varying the substrate thickness. This staggered potential medium can be realized using an appropriate nanostructured substrate for bilayer graphene (BLG), similar to the approaches used for monolayer graphene deposited on SiC and hexagonal boron nitride (h-BN) substrates \cite{wang2016gaps,nevius2015semiconducting,giovannetti2007substrate,zhou2007substrate,schnez2008analytic}. This methodology has been successfully used to simulate electronic confinement in BLG nanostructures \cite{zarenia2010simplified}.

Moreover, the resulting phase diagram exhibits a remarkably rich structure when $\Delta \ge 100$ meV.
For example, when a staggered potential medium of $\Delta = 200\text{ meV}$ is introduced (see Fig.~\ref{fig3}(c)), the band gaps for both valleys remain open, preserving a finite global bulk band gap and leading to a valley-polarized Chern insulating state with $C = -1$ ($C_K = -3/2$, $C_{K'} = +1/2$), characterized by an asymmetric distribution of Berry curvature between the two valleys—realizing a photo-induced Chern state with valley-asymmetric topology.

Additionally, Fig.\ref{fig3}(d) shows that, when one layer interacted with a staggered-potential medium with $\Delta =100$ meV,the gap closed at only one valley. This feature highlighted the presence of a single-valley Dirac-cone state: the gap remained open at the $K$ point, where the dispersion was parabolic, whereas it closed at the K$'$ point, where the dispersion is linear. To realize this interesting case, both time-reversal and spatial-inversion symmetries had to be broken, with only one layer being polarized. The SiC substrate broke inversion symmetry, whereas the circularly polarized light broke time-reversal symmetry. When both symmetries were broken, the gaps at the  K and K$'$  points became different. This configuration allowed single-valley propagating states, providing a signature of the PVQHS observed for $\Delta =100$ meV  and $V=0$.  

When the bias voltage was set to 70 meV the overall band gap closed (see Fig. \ref{fig3}(e)), leaving only valley-resolved gaps. Importantly, reversing the direction of the electric field switched the polarization of the two innermost bands, making this configuration suitable for valley-filtering applications. These valley-resolved gaps produced valley-insulating behavior by suppressing the thermal excitation of carriers between the highest occupied valence band and the lowest unoccupied conduction band, as these bands originated from different valleys.

\begin{figure}[tp]
\centering
\includegraphics[width=8cm, height=9cm]{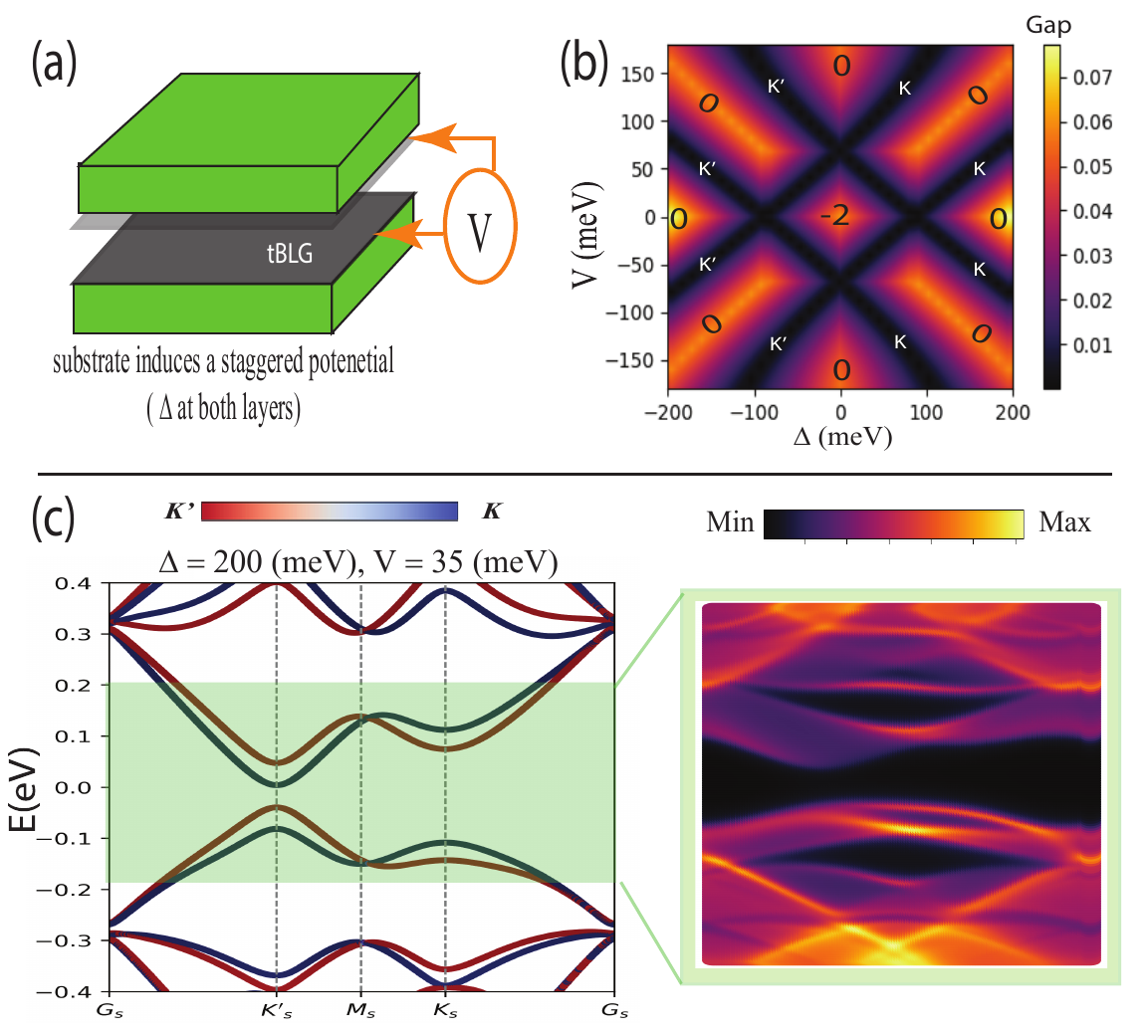}
\vspace{-0.15cm}
\caption{In panel (a), we show the sketch of the adopted tBLG-based heterostructure
 and we present the bulk band gap and the corresponding Chern number as a function of the external tuning parameters, in the presence of irradiation. In (b) we plot the  valley-resolved band structure and its surface spectral function for  $\Delta=200$~meV and  $V=35$~mV. 
 }
\label{fig4}
\end{figure}

Based on the preceding discussion, simultaneously breaking spatial-inversion symmetry in only one layer and time-reversal symmetry lifted the valley degeneracy, resulting in quantum anomalous Hall states characterized by different Chern numbers. In what followed, we investigated the case in which a staggered potential was applied to both layers, as depicted in Fig. \ref{fig4}(c).

To investigate the effect of asymmetry, we considered a surrounding medium—like the SiC substrate—that induces a staggered potential at both layers (SiC-tBLG-SiC). As shown in Figure~\ref{fig4}, when symmetry breaking is present in both layers \cite{zhai2020spin,xavier2016electronic,da2014geometry,schnez2008analytic}, the CPL interacts with the system and acts as a momentum-space perturbation that breaks time-reversal symmetry and affects the two valleys ($K$ and $K'$) in opposite ways by reversing the sign of their effective masses. In such a situation, where both layers sustain staggered potentials, the Berry curvature contributions from the two layers cancel each other, leading to a vanishing total Berry curvature. Consequently, the band structure presents a trivial gap with a Chern number $C=0$, as depicted in the phase diagram in Figure~\ref{fig4}(b) and shown in the band structure of Figure~\ref{fig4}(c). 
 
The ability to realize topological states, either in the framework of  QAHS or PVQHS, depending solely on the layer in which symmetry breaking is induced (e.g., $C=−1$ for tBLG/SiC and $C=0$ for SiC/tBLG/SiC), is particularly interesting. This tunability enables the engineering of interface states at junctions or domain walls separating regions with distinct topological invariants. Such interfaces are expected to host robust, valley-polarized channels, as discussed in detail in Section \ref{subsec3}.

\begin{figure*}[tp]
\centering
\includegraphics[width=7cm, height=5.5cm]{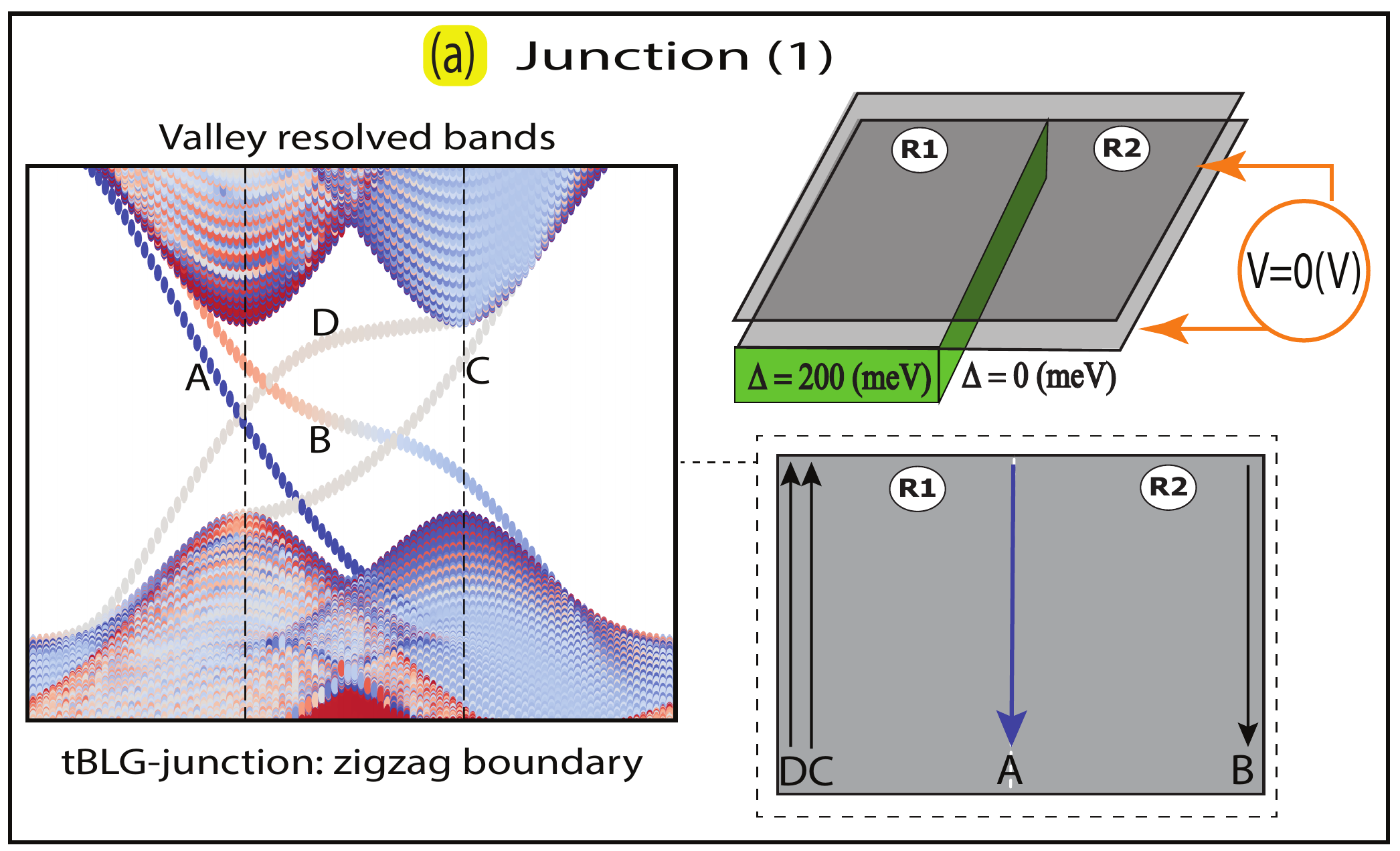}
\includegraphics[width=7cm, height=5.5cm]{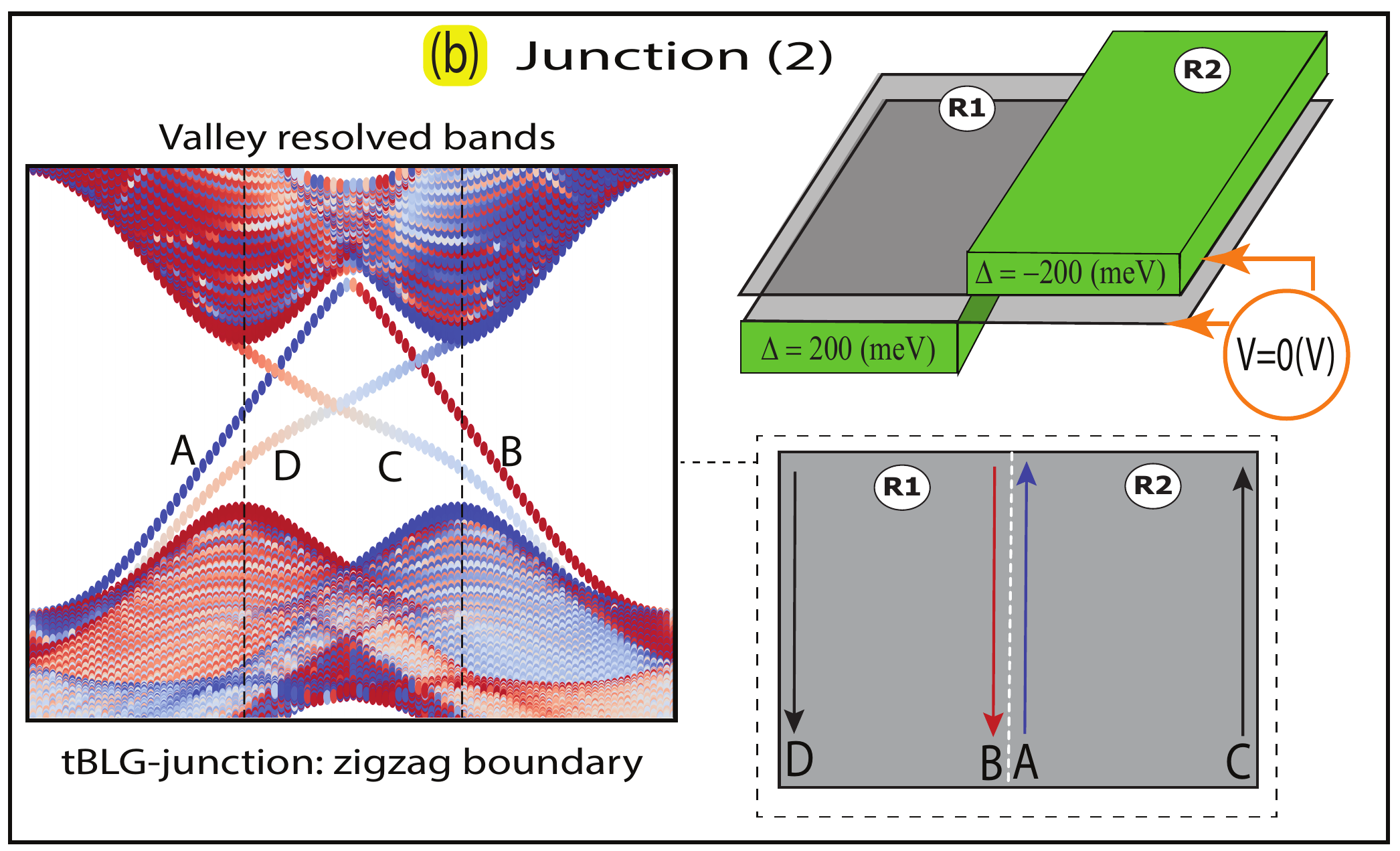}
\includegraphics[width=7cm, height=5.5cm]{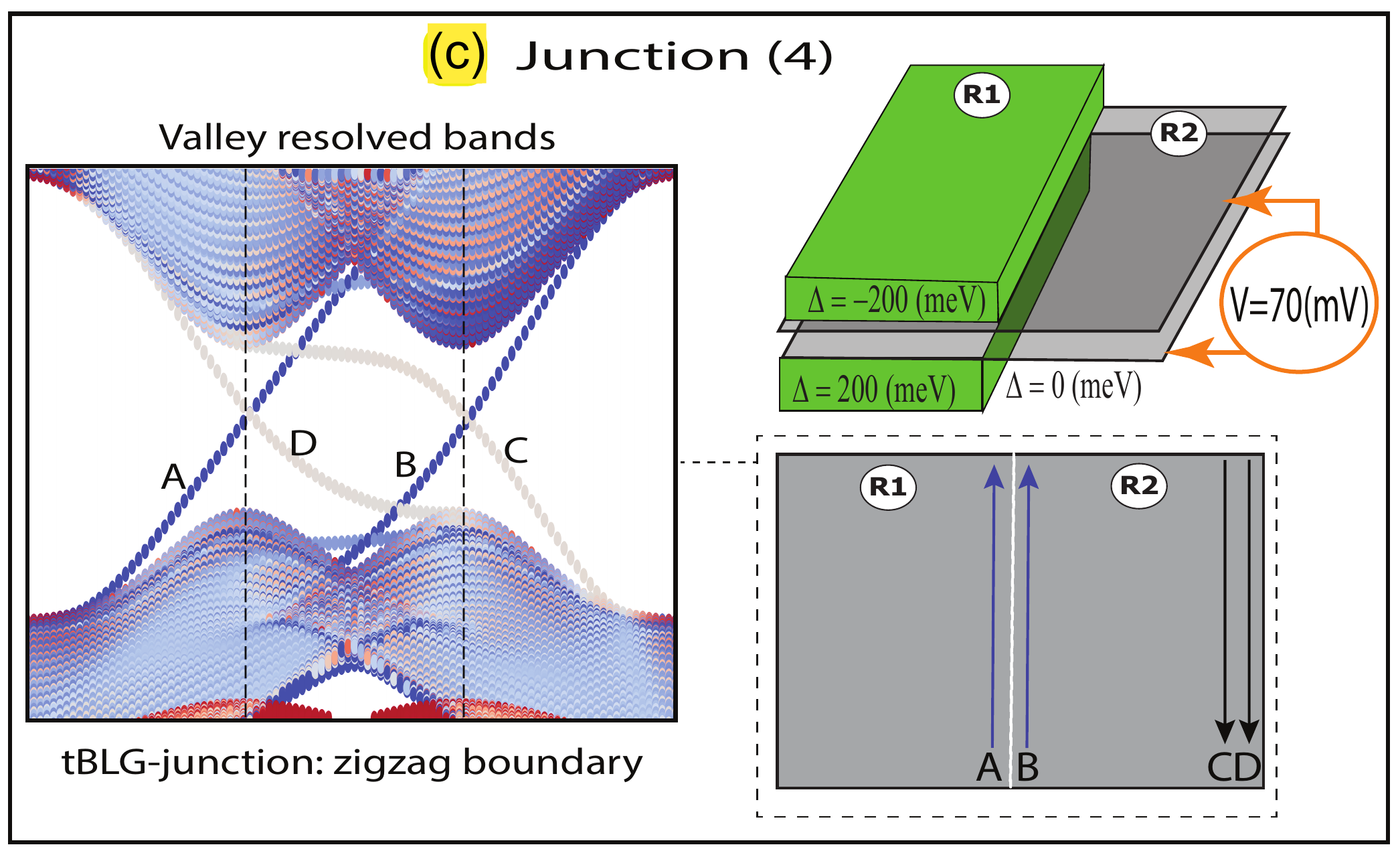}
\includegraphics[width=7cm, height=5.5cm]{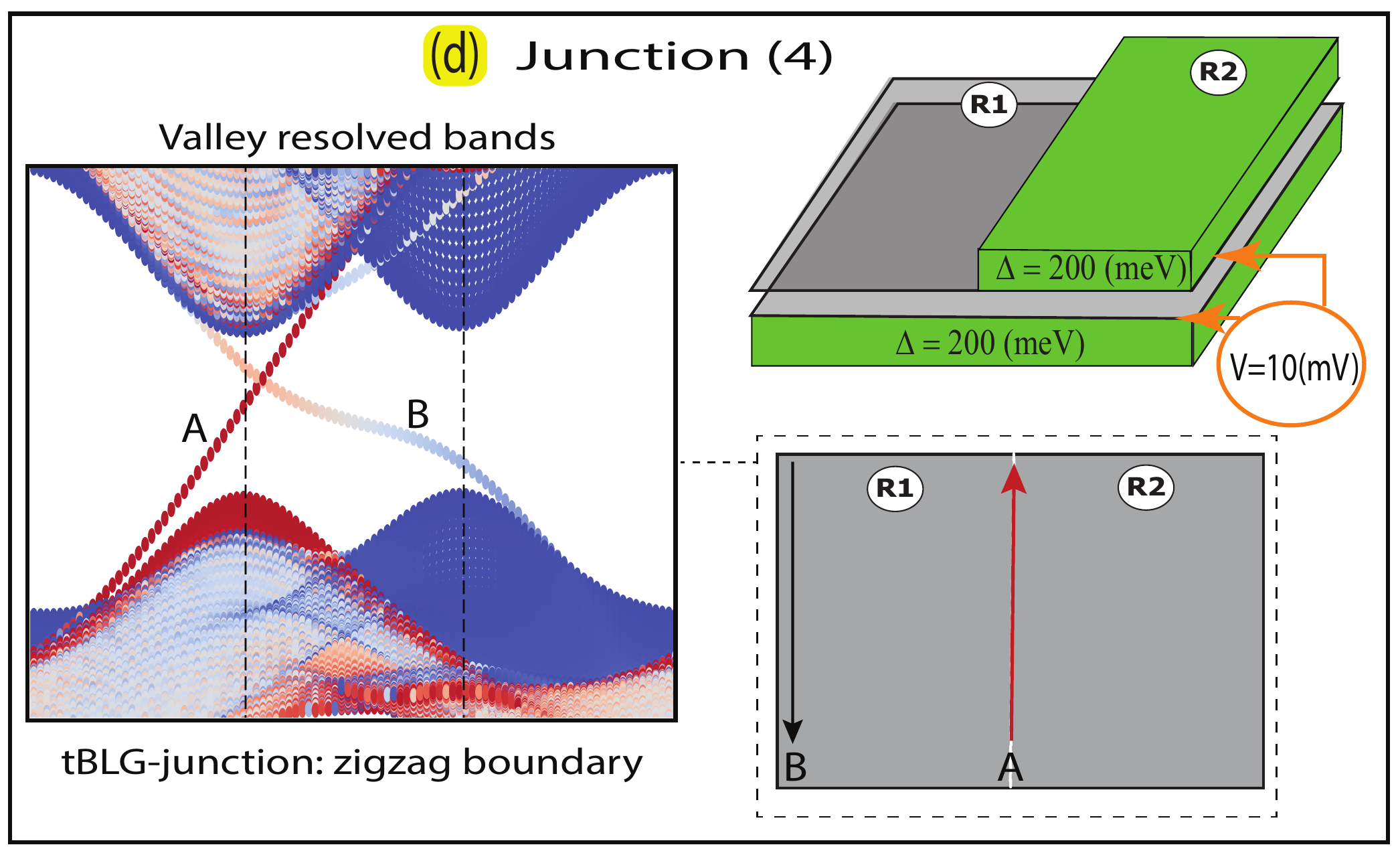}
\vspace{-0.15cm}
\caption{
Valley resolved zigzag band structures, schematic junctions, and spatial profiles of edge and interface states. Panels correspond to (a) tBLG–SiC $\mid$ tBLG, (b) tBLG–SiC $\mid$ SiC–tBLG, (c) SiC–tBLG–SiC $\mid$ tBLG, and (d) tBLG–SiC $\mid$ SiC–tBLG–SiC junctions. Red (blue) arrows indicate perfectly valley-polarized currents in the K (K') valley, while black arrows denote the presence of both valley states. These interfaces are expected to act as perfectly valley-polarized channels supporting QVHS and QHS.
}
\label{fig5}
\end{figure*}

\subsection{Domain-Wall or junction-based valley-polarized channels}\label{subsec3}
We proposed a possible realization of domain-wall-based valley-polarized channels in tBLG heterostructures. Such systems were expected to host purely valley-polarized interface states. Based on the topological-phase analysis presented in Figs. 3 and 4, topological interface states were expected to emerge whenever the topological indices differed across the interface. Consequently, tBLG-SiC with different bias configurations and/or  SiC-tBLG-SiC interface is expected to serve as a perfect valley-polarized channel host. In the following, we investigate several representative interface configurations and analyze the corresponding topological interface states. (see Fig.\ref{fig5}).

For the tBLG–SiC ($C=−1$) $\mid$ tBLG ($C=−2$) junction shown in Fig.\ref{fig5}(a), the band structure shows four interface modes within the bulk gap, labeled A–D. Among these, only mode A is purely valley-polarized, while the remaining modes are valley-mixed and confined to the outer edges (left and right boundary), propagating in opposite directions, as indicated by the black arrows. The interface mode is pinned to the bulk gap and remains throughout the bulk gap Fermi-energy range 
reflecting the nontrivial topological mismatch between the two regions. 
In Fig.\ref{fig5} (b), corresponding to a tBLG–SiC  $\mid$ SiC–tBLG junction, the situation changes qualitatively. Although the indices on the two sides are equal in magnitude, the reversal of the layer asymmetry switches the valley-resolved Chern contributions between both valleys. In case of positive staggered energy, we obtain ($C_K=-3/2$) and ($C_{K'}=+1/2$), whereas for negative staggered energy these contributions are interchanged, ($C_K=-1/2$) and ($C_{K'}=+3/2$). Thus, the opposite changes in the valley-resolved contributions across a positive-to-negative staggered-potential interface, lead to counter propagating valley-polarized interface channels. The spatial separation between the polarized interface channel and the non-polarized edge modes suppresses valley mixing and resulted in more robust valley-selective transport than that shown in Fig.\ref{fig5}(a).


A different topology is realized in Fig.\ref{fig5}(c) for the SiC–tBLG–SiC (0) $\mid$ tBLG (−2) junction. Here, two interface modes (A and B) appear within the bulk gap, both propagating along the interface and polarized in the same valley, while the outer edges host valley-degenerate modes.  Finally, in Fig.\ref{fig5}(d), we have tBLG–SiC (−1) $\mid$ SiC–tBLG–SiC (0). The band structure shows the emergence of valley-polarized channel that persists along the interface (red arrow), showing robustness against the shift in Fermi level within the bulk gap. We concluded that only state A in the band structure was purely valley-polarized.

\begin{figure*}[tp]
\centering
\includegraphics[width=16cm, height=20cm]{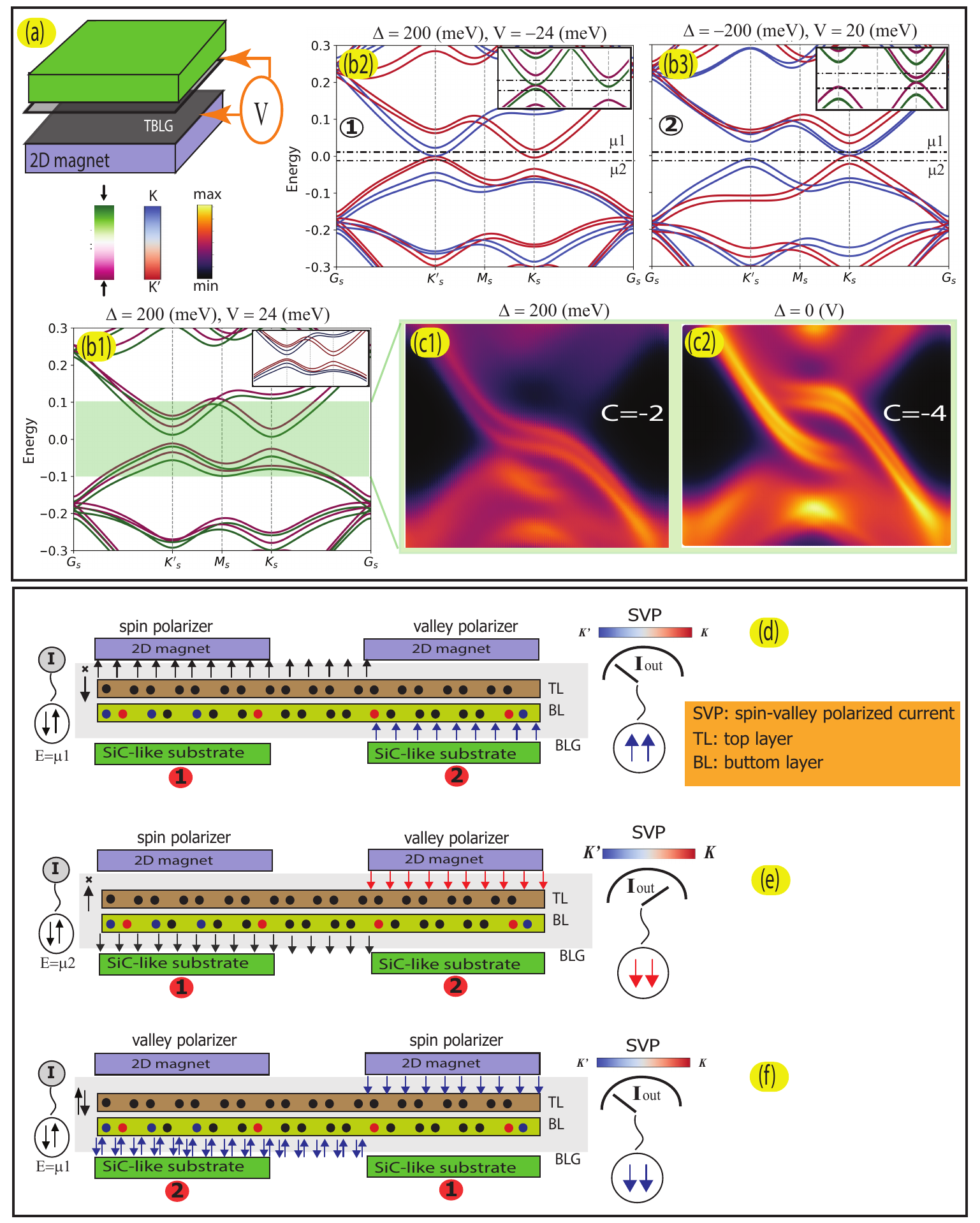}
\vspace{-0.15cm}
\caption{ we plot the valley or spin resolved band structure in (b1-b3) for tBLG-based heterostructure encapsulated by SiC substrate or ferromagnetic contact which induce, as shown in panel (a). In panels (vc1 and c2), we show the surface spectral function for $\Delta=200$ meV and $\Delta=0$meV. In panels (d, e and f) we show a schematics of a spin-valley filter device, consisting of tBLG on a substrate with two regions of  SiC and ferromagnetic substrate with different bias voltage.  
 }
\label{fig6}
\end{figure*}

\subsection{Spin-valley-Polarized Channels with Magnetic Asymmetry in tBLG-Heterostructure}

Ferromagnetic insulators (FMIs), such as EuO and Yttrium-Iron-Garnet (YIG) long been established as materials for inducing magnetic effects in graphene-based spintronics. More recently, transition-metal trihalides, such as (CrI$_3$) and Transition-Metal Trichalcogenides, such as (Cr$_{2}$Ge$_{2}$Te$_6$ or CGT), including their monolayer forms, have emerged as promising materials for low-dimensional spintronics because of their intrinsic ferromagnetic insulating ground states. The interplay between the strong proximity-induced exchange coupling provided by materials such as CGT and the substrate-induced staggered potential in SiC-supported bilayer graphene presented an important opportunity for controlling spin and valley degrees of freedom. Therefore, this section investigated the electronic structure and spin–valley-dependent transport properties of SiC–BLG–CGT heterostructures to assess the tunability achieved by combining these staggered-potential and magnetic effects.

We investigated how the SiC–BLG–CGT heterostructure modified the electronic, valley, and spin properties of bilayer graphene through proximity-induced exchange coupling, thereby establishing an important framework for interpreting future low-temperature magnetotransport and spin-pumping experiments.

Fig. \ref{fig6} (b1-3) display the calculated  electronic band structure of the twisted bilayer graphene  sandwiched between a  CGT substrate and SiC, shown along the high-symmetry lines. The combined effects of the proximity-induced exchange coupling and the staggered potential are critical: they lift both valley and spin degeneracy and {open a band gap at specific bias voltages. This calculated band profile closely resembles the bands shown in Fig. \ref{fig3} (c). Furthermore, the presence of the magnetic medium influences the topological properties, resulting in a switching the Chern number $C=−4$ for the tBLG/CGT structure, Fig. \ref{fig6} (c1), and $C=−2$ for the SiC-tBLG-CGT structure, see Fig. \ref{fig6} (c2).

The presence of an out-of-plane electric field tunes the system gap and switches between the different topological phases. Now, if we apply a positive external electric field, it adds to the internal field, opening the orbital gap further as in Fig.~\ref{fig6}(b1). But when the applied field is negative and strong enough to compensate the built-in electric field, the orbital gap closes, see Fig.~\ref{fig6}(b2), keeping the band-dependent exchange splitting unchanged. A further reversal of the staggered potential via a strong negative electric field leads to an inversion of the band ordering (see Fig.~\ref{fig6}(b3)). This reversal in band polarization occurs because the total electric field (external plus built-in field) is now pointing from the tBLG towards the SiC substrate. This mechanism achieves an exchange valley valve state. Separately, switching the magnetization of the ferromagnetic substrate would instead result in a proximity exchange spin valve state.

Designing devices based on spin–valley manipulation necessitated the ability to control both spin and valley degrees of freedom efficiently through electrical means. Our heterostructures were specifically designed to provide this high degree of electrical tunability.

The proposed spin-valley filter structure is illustrated in Fig. \ref{fig6} (d, e and f), consisting of twisted bilayer graphene deposited on a SiC-like substrate. CGT is placed on top of the tBLG, though other FMIs (YIG, EuO, CrI3) are also viable candidates. Bias voltages ($V_1$=24 mV and $V_2=−2$~ mV) are applied to the regions covered by the FMIs, which maintains the same magnetization direction.

An unpolarized current (valley and spin) is injected from the left and enters the first heterostructure region (R1: Spin Filtering). Assuming the gate voltage $V_1$ is positive, the low-energy conduction bands in this region are formed in the upper-activated layer, exhibiting valley and spin splitting (bands shown in Fig. \ref{fig6} (b1, b2)). If the chemical potential, $E_F=\mu_1 \ge 0$ (Fig. \ref{fig6} (d)), is positioned between the two spin-split electron bands, both valley channels remain open, while one spin component is filtered (see inset, Fig. \ref{fig6} (b2)). (2) In the central, bare tBLG section, the spin and valley polarizations are preserved, as the intrinsic spin-orbit coupling (SOC) is neglected. (3) Region R2 (Valley Filtering): The current then enters the second heterostructure (R2). The state of this region can be either ON or OFF depending on whether spin or valley is considered. Since the magnetization of the second FMI is the same as the first, the resulting band structure is described by Fig. \ref{fig6} (b3). In this configuration, both spin channels remain open (see inset, Fig. \ref{fig6} (b3)), while only one valley component is filtered. However, if  the chemical potential, $E_F=\mu_2 \le 0$  (Fig. \ref{fig6} (e)), the entire process is reversed. Additionally, switching the regions reversed both the spin and valley polarization of the injected current, as depicted in Fig. \ref{fig6} (b3). Consequently, the entire system functions as a spin-valley polarized channel.



\section{Conclusions}

In this work, we presented a comprehensive investigation of irradiated twisted bilayer graphene (tBLG) heterostructures. The systems considered were coupled to SiC-like substrates and ferromagnetic contacts, which generated distinct proximity effects: substrate-induced sublattice asymmetry and exchange interactions within the active tBLG layers. We examined a range of geometrical and material configurations and demonstrated how external perturbations broke spatial-inversion and time-reversal symmetries, thereby modifying the system’s electronic structure and emergent topological properties.

Our results demonstrated a highly versatile valleytronic platform whose performance depended strongly on the chosen heterostructure architecture. Through systematic analysis, we showed that these systems provided a robust foundation for tunable topological electronics. In particular, modulating the vertical electric field dynamically controlled the band topology and enabled transitions between states with distinct Chern numbers, specifically $C=-2$, $C=-1$, and 0. These electrically driven topological transitions could be directly exploited in switching-device applications, in which different topological phases corresponded to distinct conducting states, thereby enabling low-power, reconfigurable electronic and valleytronic devices.

A key outcome of this study is the interplay between circularly polarized irradiation and inversion-symmetry breaking induced by the SiC-like substrate, which enabled the stabilization of a photoinduced valley-polarized quantum Hall state. In this regime, we uncovered a distinctive single-valley Dirac phase, where one valley remained gapless with linear dispersion (active), while the other was gapped and exhibited a parabolic dispersion (inactive), effectively lifting the valley degeneracy.

Beyond bulk properties, we also explored the potential of interface and channel engineering. In particular, we showed that junctions between different topological regions (such as tBLG/SiC vs. SiC/tBLG/SiC) resulted in valley-polarized channels localized at the domain walls. These interface states exhibited remarkable robustness against perturbations, remaining largely insensitive to moderate variations in the Fermi level.

Furthermore, we extended the functionality of the proposed platform to include spin–valley filtering. By incorporating magnetic proximity effects, we designed an independent electrical control of both spin and valley degrees of freedom. This architecture effectively operates as an \textit{exchange spin-valley valve}, representing a key building block for high-density and multifunctional information processing. 

In summary, this study provided a comprehensive proof-of-concept for integrating valleytronic and spintronic functionalities within tBLG-based two-dimensional heterostructures. We identified several underlying physical mechanisms, including Floquet engineering and proximity-induced exchange interactions which enabled precise control of valley degrees of freedom and the stabilization of multiple topological phases.


\bibliography{ref.bib}
\end{document}